\documentclass[%
 aip,
 amsmath,amssymb,
reprint,%
]{revtex4-1}

\usepackage{graphicx}
\usepackage{dcolumn}
\usepackage{bm}

\usepackage[utf8]{inputenc}
\usepackage[T1]{fontenc}
\usepackage{mathptmx}
\usepackage{etoolbox}
\usepackage[dvipsnames]{xcolor}
\usepackage{framed}

\colorlet{shadecolor}{RubineRed!15}

\makeatletter
\def\@email#1#2{%
 \endgroup
 \patchcmd{\titleblock@produce}
  {\frontmatter@RRAPformat}
  {\frontmatter@RRAPformat{\produce@RRAP{*#1\href{mailto:#2}{#2}}}\frontmatter@RRAPformat}
  {}{}
}%
\makeatother
\begin{document}

\title{Basin entropy as an indicator of a bifurcation in a time-delayed system}
\author{Juan Pedro Tarigo}
\affiliation{ 
Instituto de Física, Facultad de Ciencias, Universidad de la República, Igua 4225, 11400 Montevideo, Uruguay
}%
\author{Cecilia Stari}%
\affiliation{ 
Instituto de Física, Facultad de Ingeniería, Universidad de la República, Julio Herrera y Reissig 565, 11300 Montevideo, Uruguay
}%
\author{Cristina Masoller}
\affiliation{%
Departament de Fisica, Universitat Politècnica de Catalunya, Rambla Sant Nebridi 22, Terrassa 08222, Barcelona, Spain.
}%
\author{Arturo C. Mart\'i}
\email{marti@fisica.edu.uy}

 \homepage{http://www.fisica.edu.uy/~marti}
\affiliation{ 
Instituto de Física, Facultad de Ciencias, Universidad de la República, Igua 4225, 11400 Montevideo, Uruguay
}%
\date{\today}

\begin{abstract}
The basin entropy is a measure that quantifies, in a system that has two or more attractors, the predictability of a final state, as a function of the initial conditions. 
While the basin entropy has been demonstrated on a variety of multistable dynamical systems, to the best of our knowledge, it has not yet been tested in systems with a time delay, whose phase space is infinite dimensional because the initial conditions are functions defined in a time interval $[-\tau,0]$, where $\tau$ is the delay time.
Here we consider a simple time delayed system consisting of a bistable system with a linear delayed feedback term. We show that the basin entropy captures relevant properties of the basins of attraction of the two coexisting attractors. Moreover, we show that the basin entropy can give an indication of the proximity of a Hopf bifurcation, but fails to capture the proximity of a pitchfork bifurcation. Our results suggest that the basin entropy can yield useful insights into the long-term predictability of time delayed systems, which often have coexisting attractors. 
\end{abstract}

\maketitle

\textbf{Quantifying the predictability of the long-term evolution of nonlinear dynamical systems that have coexisting attractors is an open challenge
because it is not possible to determine, in general, towards which attractor the system will evolve to, from a given initial condition.  
Time-delayed systems pose the additional challenge of an infinite-dimensional phase space.
The basin entropy is a tool that has been proposed to quantify, in a multistable system, the predictability of a final state.
Here we test the suitability of the basin entropy concept for quantifying the predictability of the long term evolution of a simple time delayed system with two coexisting fixed points. We show that the basin entropy captures the complexity of the basins of attraction of the fixed points, represented by functions with two parameters that define the initial conditions in the time interval $[-\tau,0]$, where $\tau$ is the delay time. We also show that the basin entropy gives an indication of the proximity of a Hopf bifurcation, but is not affected by the proximity of a saddle-node bifurcation.
}

\section{Introduction}

The emergence of artificial intelligence (AI) and associated technologies has brought the problem of predictability in real systems to the center of scientific debate. 
While AI models have demonstrated unprecedented predictive ability, their contribution to improving our understanding of the fundamental principles underlying complex behavior is limited \cite{sanjuan2021artificial}.
Indeed, recent work\cite{lai_2020,gauthier_2022,parlitz_2023} has shown the exceptional predictive ability of AI algorithms, even in high-dimensional chaotic systems. 
In chaotic systems, increasing our prediction horizon by one unit by using standard numerical simulation techniques implies multiplying by a factor the computation time required, making long-term predictions practically unfeasible \cite{sauer1997long}. 
However, AI algorithms trained on the basis of solutions obtained by traditional simulation methods are capable of increasing the predictive capacity by several orders of magnitude, significantly extending the predictability horizon \cite{Gelbrecht_2021}.

Chaotic systems characterized by a high sensitivity to initial conditions and by the presence of multiple attractors constitute challenging models for predictability studies, and particularly interesting are systems that display extreme multistability\cite{feudel_2015,hilda_2019,Yanchuk_2023}. On the other hand, time-delayed systems, which are ubiquitous in nature\cite{erneux2009applied}, also pose challenges for predictability studies because they can have multiple attractors that live in high-dimensional phase spaces\cite{gros_2019,radons_2019}. Such attractors, which coexist for appropriate control parameters, can be found by selecting different functions for the initial conditions \cite{masoller_1994,longtin_1997,amil2015exact}. In these systems, the basins of attraction have often riddled and intermingled structures \cite{celia_2012,amil2015organization,prasad_2016,ulrike_2018,ram_2022}. 

In a multistable system, an important problem is being able to predict, given initial conditions, to which attractor, in the long term, the system will evolve to. In practice, this question may be more relevant than forecasting the precise temporal evolution, that is, determining where, on an attractor, the system will be at a given time.

The basin entropy has been proposed for quantifying, given an initial condition, our knowledge of which attractors nearby initial conditions will evolve to\cite{daza2016basin,daza2017chaotic}. If most of the nearby initial conditions evolve towards the same attractor our uncertainty is low, and the basin entropy will be low. If, on the contrary, the nearby initial conditions evolve towards many different attractors, our uncertainty will be large and the basin entropy will also be large.

In this work our goal is to study the suitability of the basin entropy for quantifying the uncertainty with respect to which attractor the system will evolve to, in the case of a system with a time delay. We consider the simple case of a bistable system with a linear time-delayed feedback term\cite{erneux2009applied,lev,sciamanna2003optical,redmond2002bifurcation}. Since the system has, as initial condition, a function defined in the time interval $[-\tau,0]$, it has an infinite phase space.


The rest of this paper is organized as follows. Section~\ref{sec:model} describes the model analyzed, the stability of the steady state solutions and the initial function considered to integrate the equation; Sec.~\ref{sec:be} presents an overview of the basin entropy approach; Sec.~\ref{sec:results} presents the results, and Sec.~\ref{sec:conc}, our conclusions. 

\section{Model} \label{sec:model}

\begin{figure}[tb] 
\centerline{\includegraphics[width=0.9\columnwidth]{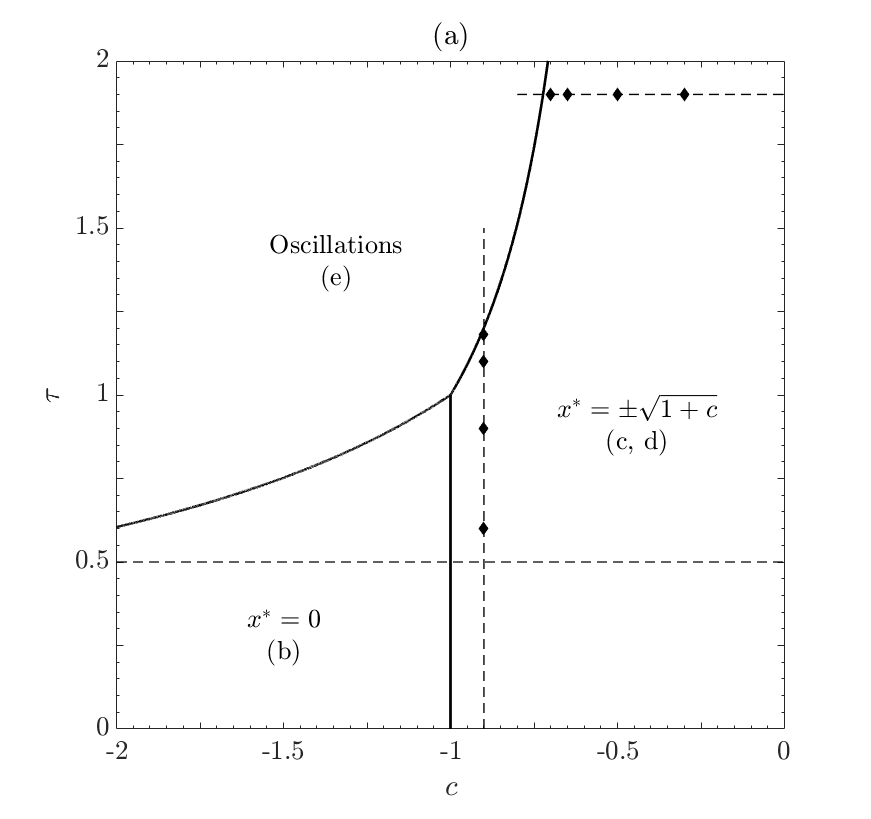}}
\centerline{\includegraphics[width=0.84\columnwidth]{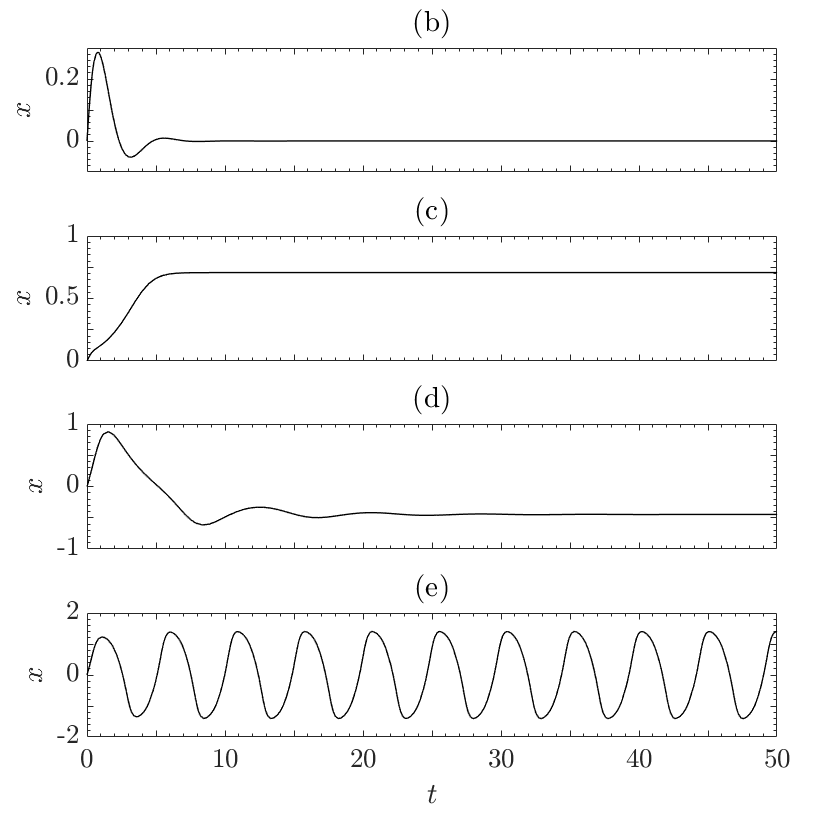}}
\caption{(a) Stability of the fixed point solutions in the parameter space ($\tau$, $c$). 
Dashed lines and diamonds indicate parameters analyzed in the following figures. 
(b-e)  Examples of transient dynamics. The parameters $(c,\tau)$ are (b) $(-1.5, 0.5)$, (c) $(-0.5, 0.5)$, (d) $(-0.7, 1.2)$, (e) $(-1.5, 1.5)$; the initial conditions are given by Eq.\eqref{eq:init} with $a = 1$ and $x_\mathrm{off} = 0$. }
\label{fig_param-space}
\end{figure}

A simple model of a bistable system with linear time delayed feedback is   \cite{erneux2009applied,sciamanna2003optical,redmond2002bifurcation}
\begin{equation}
    \dot{x} = x - x^3 + c x(t - \tau),
    \label{eq:dde}
\end{equation}
where $c$ is the strength of the feedback and $\tau$ is the feedback delay time.
Equation (\ref{eq:dde}) has three steady-steady solutions
\begin{eqnarray}
    x_0&=&0, \\
    x_\pm &=& \pm \sqrt{1+c} \quad \mathrm{for} \quad c \ge -1\end{eqnarray}
that exist independently of the value of $\tau$. The non-zero solutions appear at $c=-1$ at a pitchfork bifurcation \cite{erneux2009applied}.
The stability diagram for the steady-state solutions in the parameter space $(\tau,c)$ is presented in Fig.~\ref{fig_param-space} that also shows examples of transient dynamics. {The solid lines represent bifurcations where the stability of the solutions change. 
In the region in the top-left corner, the two fixed point solutions $x_+$ and $x_-$ coexist with oscillatory solutions \cite{erneux2009applied}. The point $(-1,1)$ corresponds to a codimension two Hopf bifurcation.}

To integrate Eq.~(\ref{eq:dde}) an arbitrary initial condition must be specified, $x_{in}(t)$, with $t$ in the interval $[-\tau, 0]$. Here, we consider as initial condition a sinusoidal function that has two parameters, the dc value, $x_{\mathrm{off}}$, and the amplitude, $a$:
\begin{equation}
    x_{in}(t) = x_{\mathrm{off}} + a \sin(t) \quad \forall t \; \in \; [-\tau, \; 0].
    \label{eq:init}
\end{equation}

\section{Basin Entropy} \label{sec:be}

We first review the definition of basin entropy of a non-delayed dynamical system that has $N_A$ coexisting attractors \cite{daza2016basin,daza2017chaotic,daza2023unpredictability}. To calculate the basin entropy we divide the phase space of the system in $N$ boxes of linear size $\epsilon$ and study towards which attractor the points in the boxes evolve, by sampling a finite number, $L$, of trajectories ($L \gg N_A$). In this way, we estimate the probability, $p_{ij}$, of ending up in attractor $j$ starting from an initial condition in box $i$ (the probabilities are normalized such that, in each box, $\sum_j^{N_A} p_{ij}=1 \forall i$). Then, the basin entropy is defined as the Shannon entropy,


\begin{equation}
S_b = - \frac{1}{N} \sum_{i=1}^N \sum_{j=1}^{N_A} p_{ij} \log p_{ij}
\end{equation}

With the normalization factor $1/N$, the basin entropy takes values between $0$ (all the initial conditions will end up in the same attractor, regarded of the box where they are in) and $\log(N_A)$ (for all the initial conditions the $N_A$ attractors have the same probability). 

To adapt this concept to a time delayed system, we consider an initial function, and analyze its parameter space; for Eq.(\ref{eq:init}), the parameter space is defined by $x_{\mathrm{off}}$ and $a$. In this space, we sample the long term evolution of $L$ trajectories ($L \gg N_A$) with initial conditions that have parameters inside a box of linear size $\epsilon$; the procedure is repeated for $N$ boxes that randomly sample the parameter space. Clearly, with this definition, the dimension of the space to be explored is the number of parameters of the initial function considered, and different initial functions can give very different values of basin entropy.
As we show in the next section, the number of boxes, $N$, has to be large enough to ensure convergence of the basin entropy, while the number of trajectories with initial parameters in each box, used to estimate the probabilities $p_{ij}$, increases with the linear size $\varepsilon$ of the box. 





\section{Results} \label{sec:results}

We integrated the model equation, Eq.(\ref{eq:dde}), with initial conditions given by Eq.(\ref{eq:init}) using a Runge-Kutta algorithm of 2nd-3rd order 
adapted to a system with delay. 
We integrated the equation up to $t = 500$ a.u. and classified the attractor reached in $N_A=4$ categories: null steady state, non-null steady states ($x_+$, $x_-$), or oscillatory behavior (not steady state). For the classification we calculated the mean value $\bar x$ and standard deviation $\sigma_x$ of the last 50 points. If $|\bar x| < 10^{-3}$ and $\sigma_x < 10^{-2}$ then the attractor is the null steady state, if $|x_\pm - \bar x| < 10^{-3}$ and $\sigma_x < 10^{-2}$ then the attractor is one of the non-null steady states and if $\sigma_x > 10^{-2}$ the attractor is not a steady state. 

\begin{figure}[tb] 
\centerline{\includegraphics[width=0.75\columnwidth]{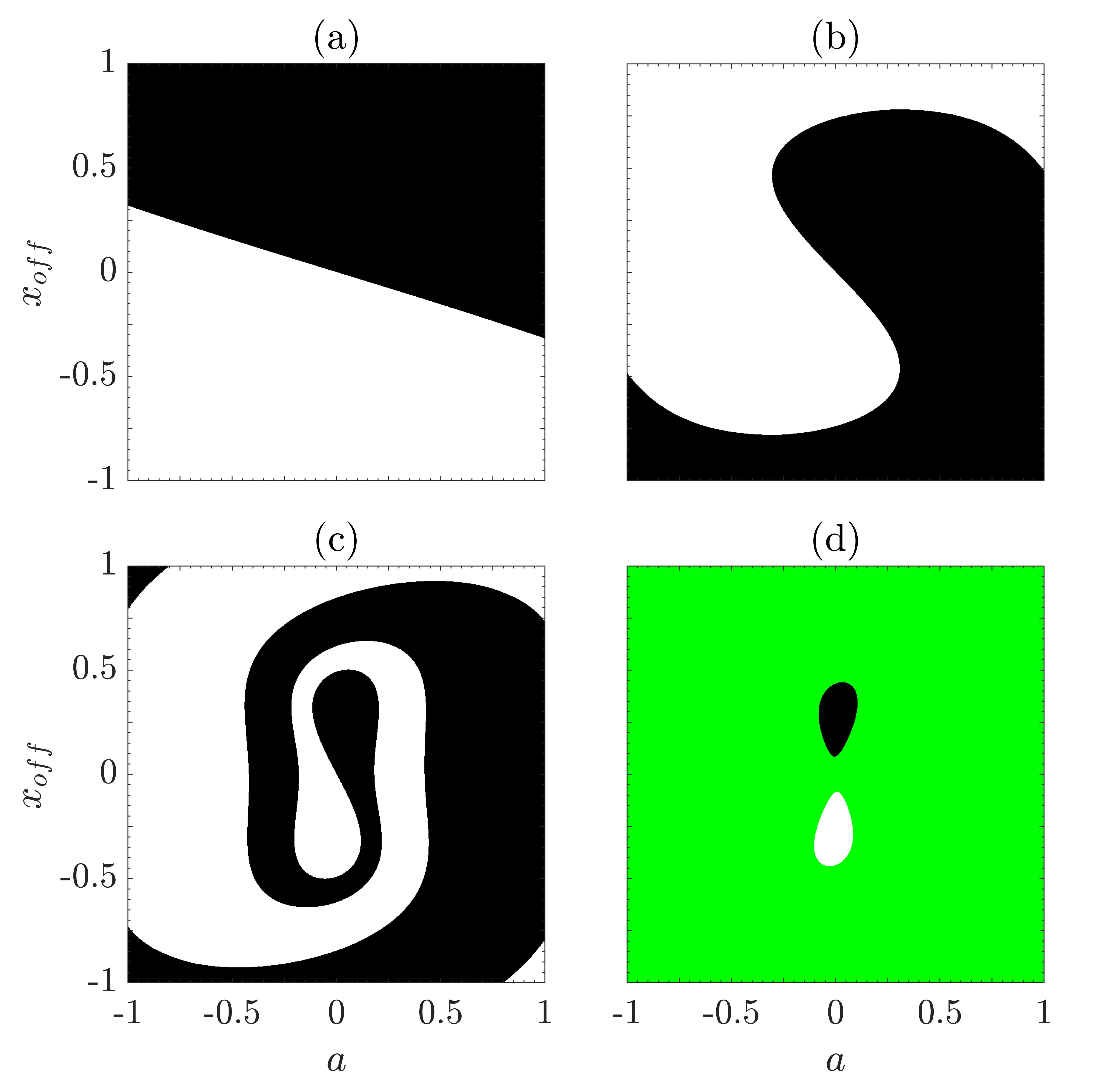}}
\caption{Basins of attraction in the plane defined by the parameters of the initial function,  $a$ and $x_\mathrm{off}$, for $c = -0.9$ and $\tau$
varying along the vertical line indicated in Fig.~\ref{fig_param-space}: (a) $\tau = 0.7$, (b) $\tau = 0.9$, (c) $\tau = 1.1$ and (d) $\tau = 1.18$. The colours indicate the type of attractor reached, green corresponds to oscillatory behavior while black and white correspond to the positive and negative fixed points respectively. Each diagram is made by simulating $1001 \times 1001$ trajectories.}
\label{fig_mapsTAU}
\end{figure}

\begin{figure}[tb] 
\centerline{\includegraphics[width=0.75\columnwidth]{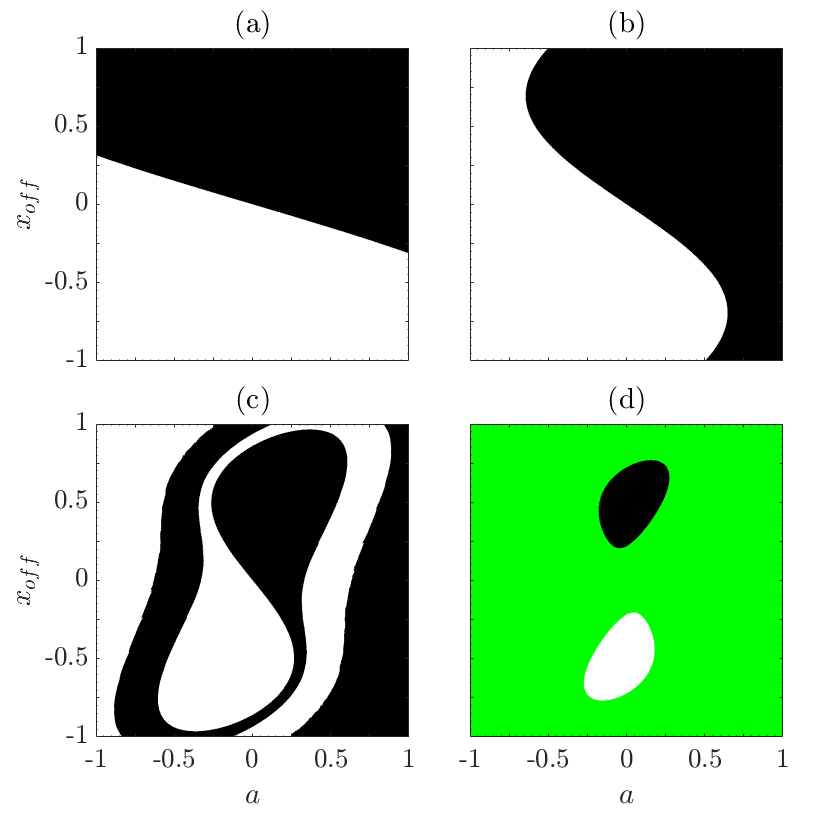}}
\caption{As Fig.~\ref{fig_mapsTAU}, but the parameter values are 
$\tau = 1.9$ and (a) $c = -0.3$, (b) $c = -0.5$, (c) $c = -0.65$ and (d) $c = -0.7$ (vary along a horizontal line in Fig.~\ref{fig_param-space}). 
}
\label{fig_mapsC}
\end{figure}


Figures~\ref{fig_mapsTAU} and \ref{fig_mapsC} depict the basins of attraction in the plane ($x_{\mathrm{off}}$, $a$); the parameters of the system, $c$ and $\tau$, take values along vertical and horizontal lines indicated in Fig.~\ref{fig_param-space}, respectively (we consider values of $c$ and $\tau$ in the region where the two non-null fixed points, $x_+$ and $x_-$, coexist). In these figures, the values of $x_{\mathrm{off}}$ and $a$ that generate trajectories that end up in $x_-$, $x_+$, or in oscillatory behavior are shown in black, white and green respectively (in the parameter region considered, the null steady state is unstable).

In Fig.~\ref{fig_mapsTAU}(a)-(c) [in Fig.~\ref{fig_mapsC}(a)-(c)] we observe that the basins of attraction of $x_-$ and $x_+$ become more intricate as $\tau$ increases [$c$ decreases]. In both cases, the system approaches the Hopf bifurcation (horizontal and vertical lines in Fig.~\ref{fig_param-space}). 
However, close to the bifurcation, Figs.~\ref{fig_mapsTAU}(d) and \ref{fig_mapsC}(d), we detected three attractors: the two fixed points, $x_-$ and $x_+$, which have small basins, and oscillatory behavior, which has a large basin. We stress that, far away from the bifurcation, we only detected the two fixed points. 

To analyze if the oscillatory behavior found close to the bifurcation may be due to critical slowing down\cite{csd} --a well-known phenomenon that causes transients to become longer and longer near bifurcations-- we analyzed the amplitude of the oscillations as a function of the integration time. We found that the oscillation amplitude first decreases and then saturates with increasing integration time, which confirms the stability of the oscillatory behavior. This is in fact consistent with a previous study that showed that, in this system, stable and oscillatory solutions might coexist \cite{redmond2002bifurcation}.

\begin{figure}[tb] 
\centerline{\includegraphics[width=0.8\columnwidth]{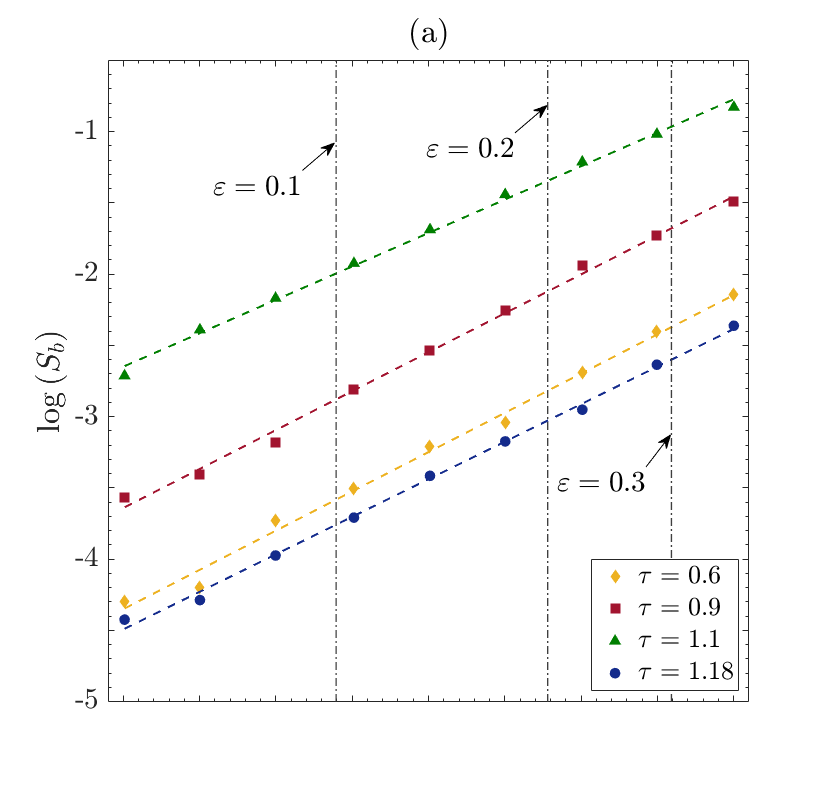}}
\centerline{\includegraphics[width=0.8\columnwidth]{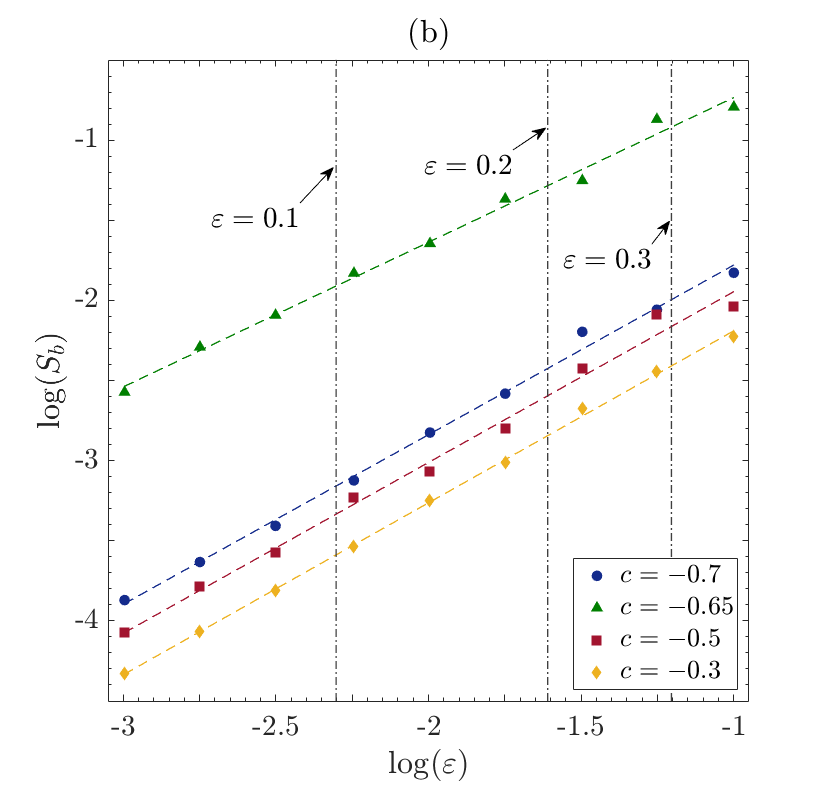}}
\caption{Log-log plot of the basin entropy as a function of the box size $\varepsilon$. The symbols show the calculated entropy while the lines display least-square fits; the slopes are equal to $1$ with less than $3\%$ uncertainties and correlation coefficient $r^2 > 0.998$. In panel (a), $c = -0.9$ is kept fixed while in (b), $\tau = 1.9$ is kept fixed. 
}
\label{fig_eps-S}
\end{figure}

Figures~\ref{fig_eps-S}(a) and \ref{fig_eps-S}(b) display the basin entropy of the basins of attraction shown in the four panels of Figs.~\ref{fig_mapsTAU} and~\ref{fig_mapsC} respectively, as a function of the box size, $\varepsilon$. In a log-log plot, the variation of the entropy with $\varepsilon$ is linear with slope very close to one, which is an indicator of  the smoothness of the boundaries between basins of attraction \cite{daza2023unpredictability}. It can also be observed that the value of the entropy is consistent with the ``complexity'' of the basin of attraction seen in Figs.~\ref{fig_mapsTAU} and \ref{fig_mapsC}:  as the bifurcation is approached the basins of attraction become  more intricate and the entropy increases (yellow, red and green symbols), but very close to the bifurcation, the structure of the basins of attraction becomes simple, Figs.~\ref{fig_mapsTAU}(d) and~\ref{fig_mapsC}(d), and the entropy decreases (blue symbols).

\begin{figure*}[tb] 
{\includegraphics[width=0.65\columnwidth]{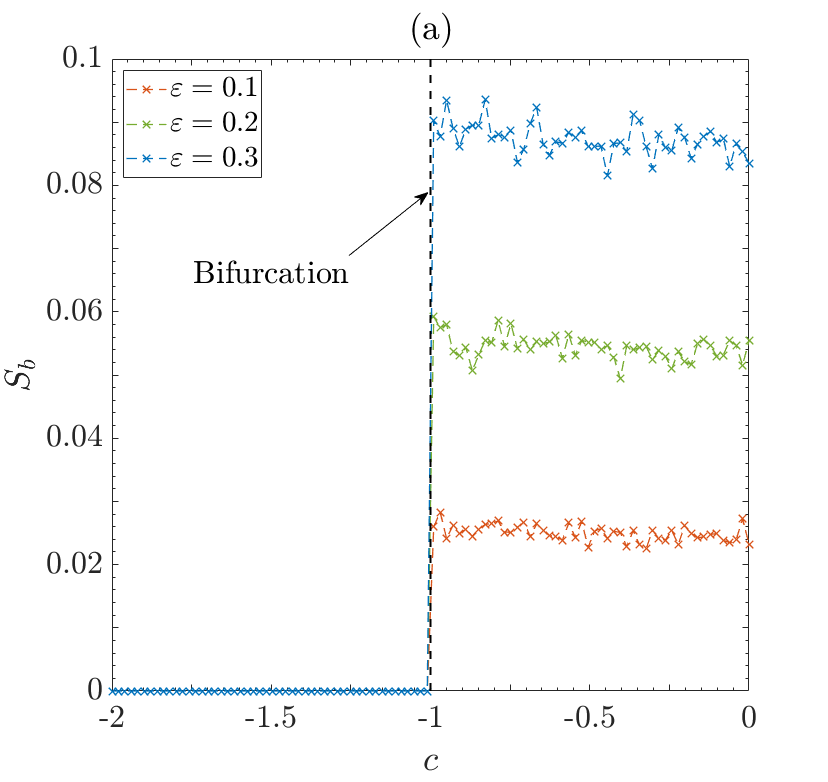}}
{\includegraphics[width=0.65\columnwidth]{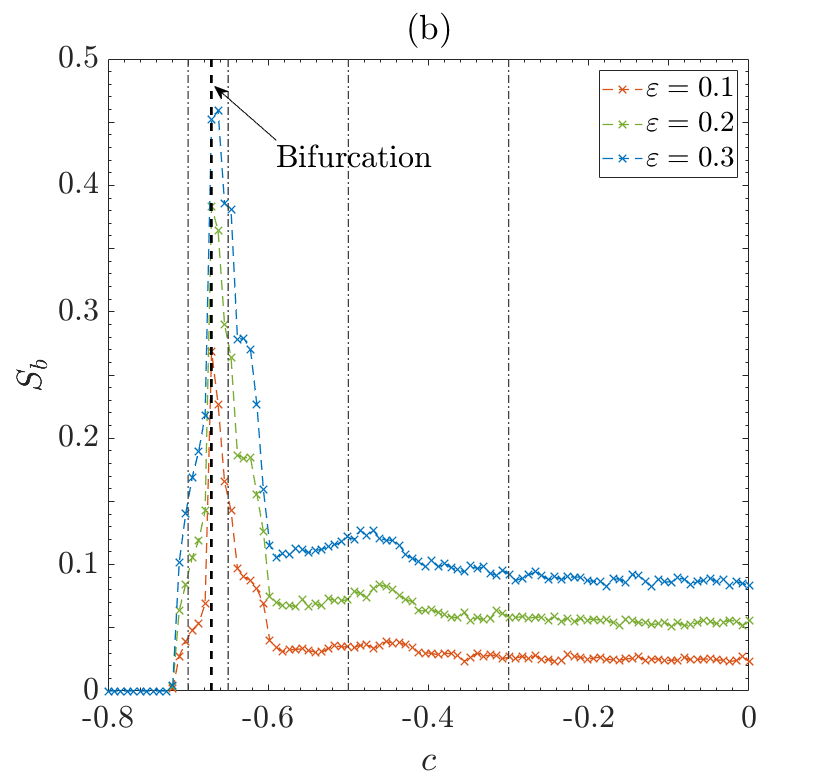}}
{\includegraphics[width=0.65\columnwidth]{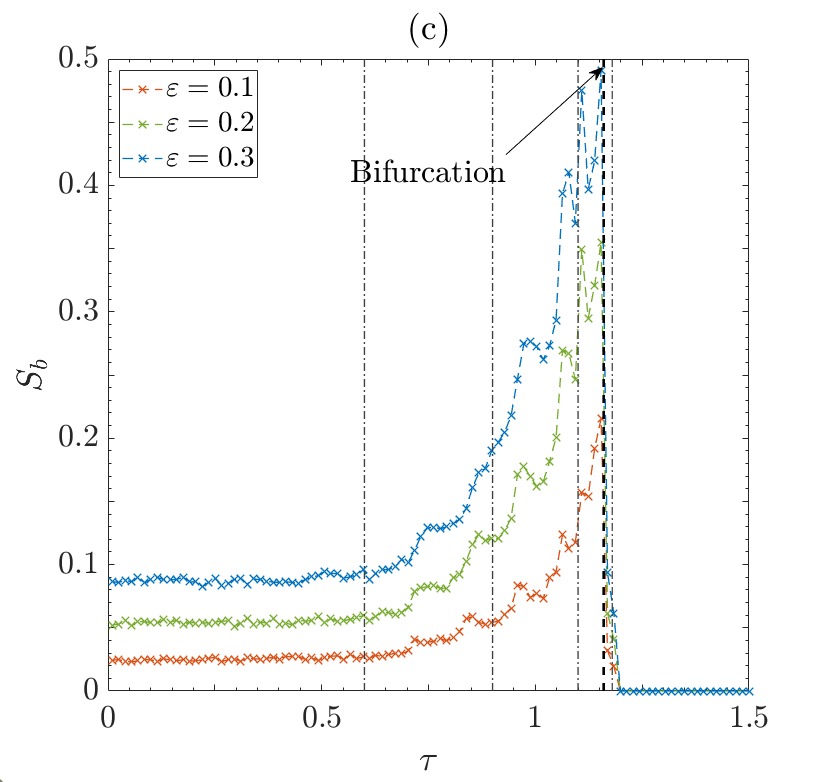}}
\caption{Basin entropy as a function of a parameter of the system. (a) $c$ varies while $\tau = 0.5$ is kept fixed, (b) $c$ varies while $\tau = 1.9$ is kept fixed,(c) $\tau$ varies while $c = -0.9$ is kept fixed. The parameters of the initial function, $a$ and $x_\mathbf{off}$, were varied between $-1$ and $1$, and the linear size of the box was $\varepsilon = 0.1$ (red line), $\varepsilon = 0.2$ (green line) and $\varepsilon = 0.3$ (blue line), and the number of trajectories with parameters of the initial function in each box was 25, 100 and 225 respectively.}
\label{fig_lines}
\end{figure*}

To analyze the evolution of the basin entropy as we approach a bifurcation, we study how it varies along horizontal and vertical lines in the parameter space ($c$, $\tau$). For this purpose, we consider boxes of linear size $\varepsilon = 0.1$, $0.2$ and $0.3$ and calculate the basin entropy as a function of $c$ or $\tau$. The results are shown in Fig.~\ref{fig_lines}. 

In Fig. \ref{fig_lines}(a), $c$ decreases while $\tau=0.5$ is kept constant. At $c = -1$ a pitchfork bifurcation occurs, and for $c \le -1$ the basin entropy is $0$ because the system has only one stable attractor, the null fixed point. When $c \ge -1$ the two non-null fixed points coexist and the basin entropy is larger than $0$; however, the basins of attraction of the two coexisting fixed points are largely unaffected when $c$ changes and therefore, the basin entropy does not show  substantial changes when the pitchfork bifurcation is approached.   



Figures \ref{fig_lines}(b) and \ref{fig_lines}(c) display the basin entropy when $c$ and $\tau$, respectively, vary and the Hopf bifurcation is approached (the parameters are the same as in Figs.~\ref{fig_mapsTAU}, \ref{fig_mapsC} and \ref{fig_eps-S}).
In both cases we see that the entropy increases as the Hopf bifurcation is approached, and reaches a maximum value just before oscillatory behavior appears, and then decreases rapidly in the region where oscillations coexist with the two fixed points. When the fixed points are unstable ($c<-0.7$ or $\tau>1.2$) the entropy is zero because only the oscillatory behavior is stable.

\begin{figure*}[tb] 
{\includegraphics[width=0.6\columnwidth]{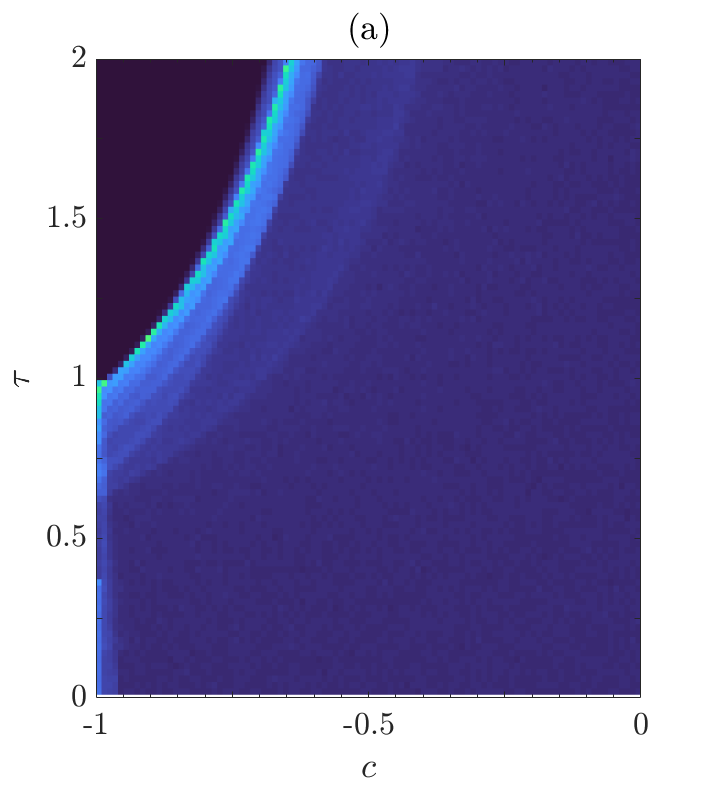}}
{\includegraphics[width=0.6\columnwidth]{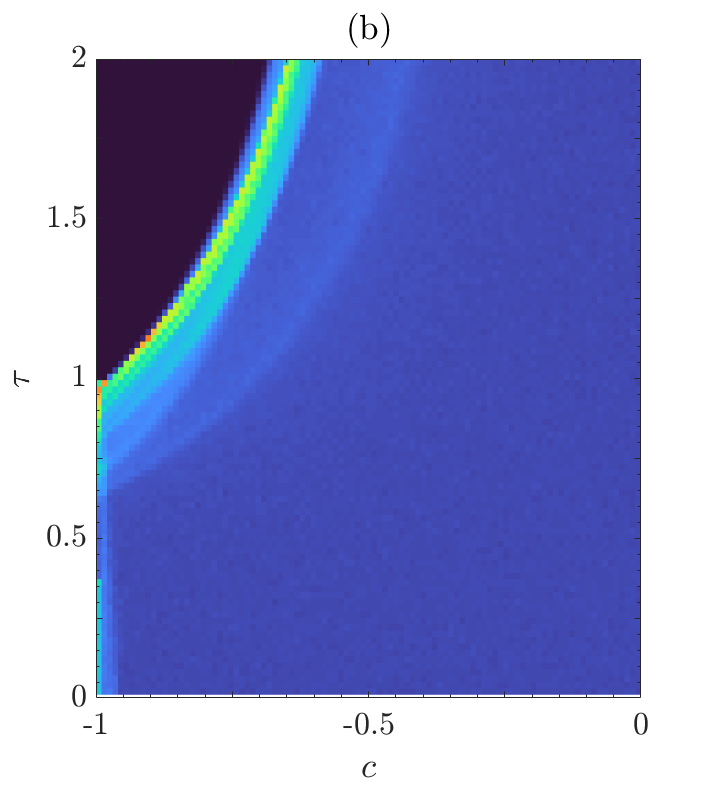}}
{\includegraphics[width=0.6\columnwidth]{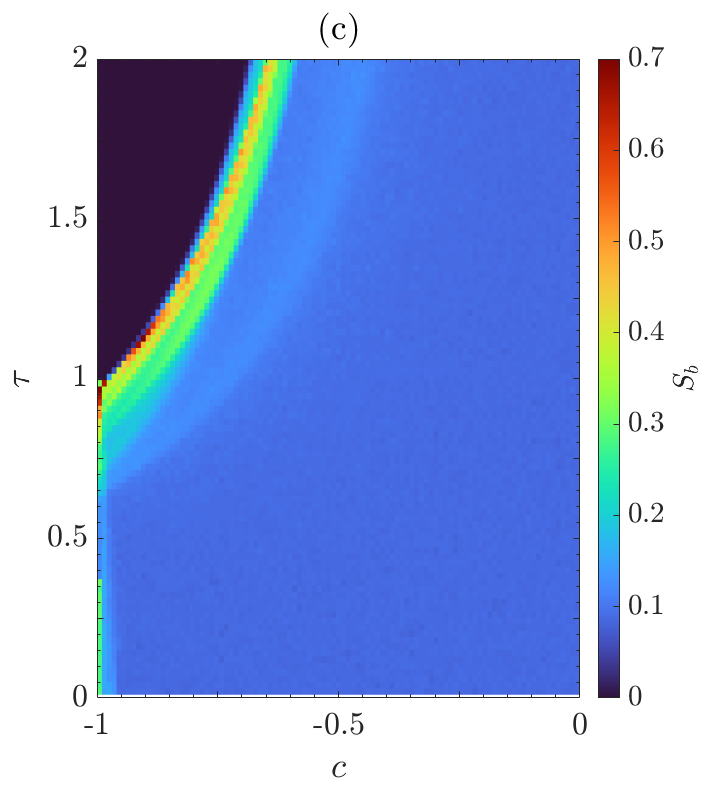}}
\caption{Basin entropy as a function of the system's parameters $c$ and $\tau$, calculated with three box sizes: (a) $\varepsilon = 0.1$, (b) $\varepsilon = 0.2$ and (c) $\varepsilon = 0.3$. Other parameters are as in Fig.~\ref{fig_lines}.}
\label{fig_BasinEntropy}
\end{figure*}

Figure~\ref{fig_BasinEntropy} summarizes the above results by displaying in color code the basin entropy, as a function of the model's parameters, $c$ and $\tau$, for the three box sizes considered. In each point or pixel the value of the basin entropy was calculated by varying the parameters $a$ and $x_\mathrm{off}$ of the initial function. We see that the entropy increases when approaching the region where the two fixed points coexist with the oscillatory behavior, and vanishes when the fixed points become unstable.


\begin{figure}[tb] 
\centerline{\includegraphics[width=0.9\columnwidth]{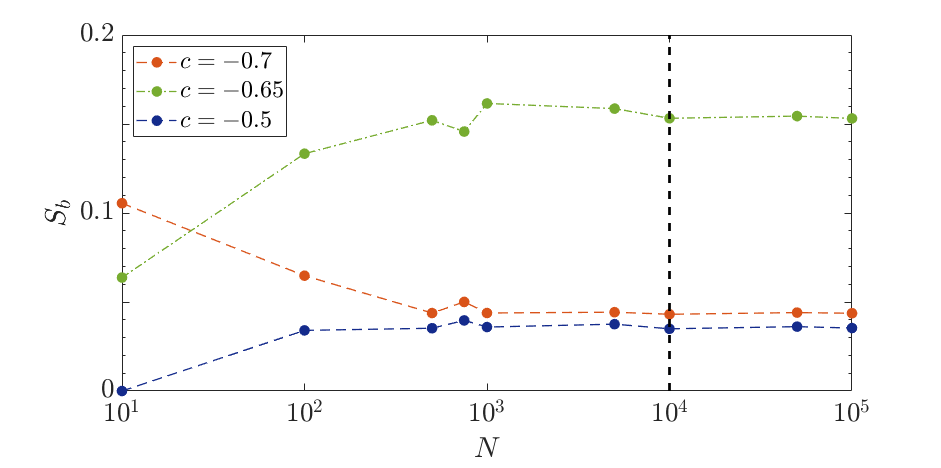}}
\caption{Basin entropy as a function of the number of boxes $N$ (log scale) for three values of $c$ and $\tau = 1.9$. Here the parameters of the initial function, $a$ and $x_\mathbf{off}$, were varied were varied in the range between $-1$ and $1$, and the box size was $\varepsilon = 0.1$.}
\label{fig_boxes}
\end{figure}

Finally, we show that these results are robust, if a sufficiently large number of boxes, $N$, is used to randomly sample the plane ($x_{\mathrm{off}}$, $a$).
Figure~\ref{fig_boxes} displays the basin entropy as a function of $N$ and we see that $N=10^4$ allows a detailed exploration of the plane ($x_{\mathrm{off}}$, $a$), while it avoids extra computational time. 

\section{Conclusion} \label{sec:conc}
We have studied the predictability of the long-term evolution of a bistable system with linear delayed feedback, using the basin entropy approach applied to a  set of initial functions in the interval $[-\tau,0]$ that have two parameters, $x_{off}$ and $a$. We have analysed the basin of attraction of four types of long-term behavior (three steady states and oscillatory behavior) in the plane defined by  $x_{off}$ and $a$. We have shown that the basin entropy can capture changes in the basins of attractions that occur when the system's parameters (the delay time or the feedback strength) vary. We have found that the entropy is a good indicator when a Hopf bifurcation is approached, but it is not able to identify the pitchfork bifurcation, because, at least for the particular system and initial function considered here, the approach to the pitchfork bifurcation does not affect the structure of the basins of attraction of the fixed points. For future work, it will be interesting to test the performance of the basin entropy with other initial functions, and to consider other time delayed systems that have other bifurcations and/or a larger number of coexisting attractors.


\begin{acknowledgments}
J.P.T., C.S. and A.C.M. acknowledge support of PEDECIBA (MEC-Udelar, Uruguay); C.M. acknowledges support of ICREA ACADEMIA, AGAUR (2021 SGR 00606) and Ministerio de Ciencia e Innovación Spain (Project No. PID2021-123994NB-C21). The numerical  experiments were performed at the ClusterUY (site: https://cluster.uy)
\end{acknowledgments}

\bibliography{ref.bib}

\end{document}